\documentclass[onecolumn,showpacs,amsmath,amssymb,prd,nofootinbib]{revtex4-2}
\usepackage{graphicx}
\usepackage{epsfig}
\usepackage{enumerate}
\def\beq{\begin{equation}}
\def\eeqno#1{\label{#1}\end{equation}}

\def\rarrow{\rightarrow }

\def\dleft{\rlap{{\it D}}\raise 8pt
\hbox{$\scriptscriptstyle\Leftarrow$}}
\def\dright{\rlap{{\it
D}}\raise 8pt\hbox{$\scriptscriptstyle\Rightarrow$}}

\def\az{a_{0}}
\def\azs{a_{0}^2}

\def\l0{\ell_{0}}

\def\rar{\rightarrow}
\def\s{\sigma}

\def\a{\alpha}
\def\b{\beta}
\def\c{\gamma}
\def\l{\lambda}

\def\f{\phi}
\def\t{\theta}

\def\r{\rho}
\def\m{\mu}
\def\n{\nu}
\def\z{\zeta}

\def\vinf{V\_{\infty}}

\def\A{\mathcal{A}}

\def\F{\mathcal{F}}
\def\L{\mathcal{L}}

\def\d{\delta}

\def\drt{d^3\vr}
\def\a{\alpha}
\def\xlimin{{x\rarrow\infty \atop{\raise 1pt\hbox to 30pt
{\rightarrowfill}}}}
\def\limlim#1#2{{#1\rarrow #2 \atop{\raise 1pt\hbox to 30pt
{\rightarrowfill}}}}
\def\eps{\epsilon}
\def\vr{{\bf r}}

\def\vv{{\bf v}}

\def\vF{{\bf F}}

\def\ve{{\bf e}}

\def\grad{\vec\nabla}
\def\div{\vec \nabla\cdot}
\def\gf{\grad\phi}

\def\L{\mathcal{L}}
\def\U{\mathcal{U}}

\def\gN{g\_N}

\def\m{\mu}
\def\a{\alpha}
\def\b{\beta}
\def\C{\Gamma}

\def\n{\nu}

\def\ten#1#2{\^{#1}\_{#2}}
\def\F{\mathcal{F}}

\def\_#1{_{\scriptscriptstyle #1}}
\def\^#1{^{\scriptscriptstyle #1}}

\def\Gmn{g\^{\mu \nu}}

\def\C{\Gamma}
\def\der#1{\_{,#1}}
\def\uder#1{\_,\^{#1}}

\def\oot{\frac{1}{2}}

\def\fpg{4\pi G}
\def\epg{8\pi G}

\def\vrf{\varphi}

\def\gps{\grad\psi}
\def\gvrf{\grad\vrf}

\begin{document}
%\eqsec  % uncomment this line to get equations numbered by (sec.num)
\title{Tripotential MOND theories}

\author{Mordehai Milgrom}
\affiliation{Department of Particle Physics and Astrophysics, Weizmann Institute}

\begin{abstract}
I present a new class of nonrelativistic, modified-gravity MOND theories. The three gravitational degrees of freedom of these ``TRIMOND'' theories are the MOND potential and two auxiliary potentials, one of which emerges as the Newtonian potential.
Their Lagrangians involve a function of three acceleration variables: the gradients of the potentials. So,  the transition from the Newtonian to the MOND regime is rather richer than in the aquadratic-Lagrangian theory (AQUAL) and the quasilinear MOND theory (QUMOND), which are special cases of TRIMOND, each defined by a Lagrangian function of a single variable. In particular, unlike AQUAL and QUMOND whose deep-MOND limit (DML) is fully dictated by the required scale invariance, here, the scale-invariant DML still requires specifying  a function of two variables. For one-dimensional (e.g., spherical) mass distributions, in all TRIMOND theories the MOND acceleration is a (theory specific, but system independent) function of the Newtonian acceleration; their variety appears in nonsymmetric situations. Also, they all make the salient, primary MOND predictions. For example, they predict the same DML virial relation as AQUAL and QUMOND, and thus the same DML $M-\sigma$ relation, and the same DML two-body force. Yet they can differ materially on secondary predictions.
Such TRIMOND theories may be the nonrelativistic limits of scalar-bimetric relativistic formulations of MOND, such as BIMOND with an added scalar.

\end{abstract}
%\keywords{}
%\pacs{04.50.Kd, 98.80.Jk}
\maketitle

\section{Introduction}
The ``aquadratic-Lagrangian'' (AQUAL) theory \cite{bm84} was the first full-fledged, nonrelativistic theory that embodies the basic tenets of MOND \cite{milgrom83}. (For reviews of MOND see Refs. \cite{fm12,milgrom14,milgrom20,mcgaugh20,merritt20,bz22}.) AQUAL modifies the Poisson equation for the gravitational potential of a given distribution of masses, without modifying the equation of motion of the masses; it is thus a ``modified-gravity'' (MG) theory. Some 25 years later another full-fledged nonrelativistic, MG Lagrangian theory for MOND has been proposed -- the quasilinear MOND theory (QUMOND) \cite{milgrom10}.
These two theories -- disparate in their amenability to solution, but yielding very similar prediction -- have become the workhorses for studying various aspects of MOND, including many detailed studies that involve numerical solutions of the field equations of these theories (see reviews  \cite{fm12,milgrom14,milgrom20,mcgaugh20,merritt20,bz22}).
\par
AQUAL and QUMOND satisfy the basic tenets of MOND -- the involvement of one new, acceleration constant, $\az$; a Newtonian limit at high accelerations ($a\gg\az$, or $\az\rar 0$); and scale invariance in the low-acceleration, deep-MOND limit (DML), $\az\rar\infty$. They thus share all the primary predictions of MOND -- those that follow from only the basic tenets \cite{milgrom14a}. Examples of such predictions are, the asymptotic flatness of rotation curves $V(r)\rar \vinf$; the relation between the total mass, $M$, and the asymptotic speed: $MG\az=\vinf^4$, (underlying the ``baryonic Tully-Fisher relation''); a good approximation to the full rotation curves from the mass distribution; enhanced stability of disc galaxies; a relation between the total mass of a system in the DML and its mass-average, root-mean-squared velocity dispersion, $\s$: $MG\az=\xi\s\^4$, with $\xi\sim 1$; and the qualitative prediction of an ``external-field effect''.
\par
Perhaps as a result of these two theories remaining the main tools for deriving quantitative results in MOND, the notion may have taken root, that they represent MOND in its entirety (in the nonrelativistic regime).
\par
It has, however, been stressed repeatedly that there may well be other theories that satisfy the basic MOND tenets, and thus share its primary predictions, but that may differ from AQUAL and QUMOND, and among themselves, in making other, secondary predictions.
Among such second-tier predictions one may count some fine details of the rotation curves (e.g., Refs. \cite{brada95,milgrom12,brown18,pl20,chae22}); the exact value of $\xi$ in the $M~-~\s$ relation, and its dependence on dimensionless attributes of the system; the exact dependence of the effective two-body force on the masses; the exact nature and strength of the external-field effect -- such as the exact effects of the galactic field on the dynamics of the Solar System or of wide binaries in the galaxy; etc.
\par
One possible avenue for looking for such other MOND formulations is that of so-called ``modified-inertia'' theories, which as described, e.g., in the recent Ref. \cite{milgrom22a}, can make very different secondary predictions than AQUAL and QUMOND.
\par
But secondary predictions can differ also among MG theories, as was demonstrated recently in  Ref. \cite{milgrom23}.
\par
AQUAL and QUMOND are quite restricted in their scope in the following sense: They each involves one ``interpolating function'' of a single acceleration variable that is introduced, by hand, in their Lagrangian. As a result, all the (secondary) predictions of these theories involve, and are dictated, by this one function.
For example, this function dictates the prediction of the ``mass-discrepancy-acceleration relation'' (MDAR also known as the RAR) and can be extracted from the observed relation. It then largely determines all the other predictions of these theories.
This is much too restrictive, and is not a situation that characterizes other, well-known modifications of Newtonian dynamics.
\par
For example, in quantum theory we do not introduce an interpolation function at the basic level. We do encounter different ``interpolating functions'' between the classical and the quantum regimes for different phenomena: The black-body function, the expression for the specific heat of solids, and the barrier penetration probability are only a few examples.
The same is true in relativity vs Newtonian dynamics.
\par
Likewise, one expects that in a more fundamental MOND theory, interpolating functions will not be put in by hand in the fundamental equations. They will be derivable from the theory, and will also differ among the different applications of the theory.
\par
In distancing ourselves at least one step from the restricted AQUAL/QUMOND, we should realize that there are MOND theories whose Newtonian-DML interpolation scheme is richer, even if it is still implanted at the basic level of the theory.
\par
Here I present a new class of nonrelativistic, MG formulations of MOND that involve three gravitational potentials, one that dictates the acceleration of masses -- and is thus the ``observed'' MOND potential -- with two auxiliary ones. The interpolating function depends on the gradients of the last two and is a function of three scalar, acceleration variables, offering a rather richer scope than AQUAL/QUMOND, which are special cases in this new class.
\par
In Sec. \ref{tripo}, I describe the TRIMOND Lagrangian and the resulting field equations for the three potentials, and consider their high- and low-acceleration limits. In Sec. \ref{conseq}, I describe some of the properties of the solutions and various predictions of the TRIMOND theories. Section \ref{discussion} is a brief discussion of potential relativistic extensions.

\section{Tripotential formulations   \label{tripo}}
The tripotential theories involve three potentials as the gravitational degrees of freedom, with one of them, the MOND potential $\f$ coupling to matter directly. The other two are called $\vrf$ and $\psi$.
The Lagrangian density of these theories is of the form
\beq \L=\L_G  +\r(\oot\vv^2-\phi),  \eeqno{lagran1}
where its gravitational part is
\beq \L_G=-\frac{1}{\epg}[2\gf\cdot\gps-\azs\F(x,y,z)] ,  \eeqno{lagra}
with the scalar, acceleration variables
\beq x\equiv(\gps)^2/\azs,~~~y\equiv(\gvrf)^2/\azs,~~~z=2\gps\cdot\gvrf/\azs, \eeqno{xyz}
and $\F$ is a dimensionless function satisfying the basic tenets of MOND (see Secs. \ref{newt} and \ref{dml}).
\par
 The density $\r$ can be viewed as made up of constituent point masses, $m_i$, moving on trajectories $\vr_i(t)$, and treatable, each, as a test mass in the mean field of the rest. Namely, $\r(\vr,t)=\sum_i m_i\d^3[\vr-\vr_i(t)]$.
The Lagrangian itself is then
\beq L=\int \L\drt,  \eeqno{mayate}
and the contribution of the second term in Eq. (\ref{lagran1}) to it is
\beq \sum_i m_i[\frac{\vv_i\^2}{2}-\phi(\vr_i)].  \eeqno{ndanfa}
\par
Variation  over the particle degrees of freedom [which appear only in expression (\ref{ndanfa})] gives
 \beq \ddot\vr_i=-\gf(\vr_i);   \eeqno{iitr}
so $\f$ is the (MOND) gravitational potential, which dictates particle accelerations.
\par
Variation over $\f$ gives
 \beq \Delta\psi=\fpg\r; \eeqno{iiatr}
so, $\psi$ equals the Newtonian potential sourced by $\r$. (As usual, we seek solutions for which the three gradients vanish at infinity for isolated systems.)
\par
Varying over $\vrf$ gives
\beq \div(\F_y\gvrf)+\div(\F_z\gps)=0.   \eeqno{vvvf}
Varying over $\psi$ gives
\beq \Delta\f=\div(\F_x\gps)+\div(\F_z\gvrf)\equiv\fpg\hat\r.   \eeqno{mail}
\par
After $\psi$ is solved for from Eq. (\ref{iiatr}), it is substituted in Eq. (\ref{vvvf}), which becomes a nonlinear, second order equation in $\vrf$, with coefficients that depend on position through $\gps(\vr)$. The choice of $\F$ should be such that Eq. (\ref{vvvf}) is elliptic, and then it has a unique solution for $\vrf$ with our boundary conditions.
Then, $\vrf$ and $\psi$ are substituted in the right-hand side of Eq. (\ref{mail}) -- a Poisson equation for $\f$ -- to get the effective density, $\hat\r$, that sources it. The reason that this successive solution scheme is possible -- i.e, that we do not have to solve coupled equations -- is that $\f$ and $\vrf$ are not coupled directly.
\par
As in AQUAL, the nonlinearity of Eq. (\ref{vvvf}) involves functions of only up to the first derivatives, the second derivatives appearing in a linear fashion with coefficients that are functions of position and the first derivatives.\footnote{Such equations are called ``quasilinear'' in the mathematical literature, to be distinguished from what we call ``quasilinear'' in describing QUMOND.}
Thus, after the substitution of $\gps$, Eq. (\ref{vvvf}) is of the form
\beq a\^{ij}(\vr,\gvrf)\vrf_{,i,j}+b(\vr,\gvrf)=0.  \eeqno{ellip}
Then, the ellipticity condition is that the matrix $a\^{ij}$ has a definite sign \cite{gilbarg77}.
\par
The solution scheme is rather simple in the case where $\F_z=\eps$ is a constant. Equations (\ref{vvvf}) and (\ref{mail}) can then be written as
\beq \div(\F_y\gvrf)=-\fpg\eps\r.   \eeqno{vvvfss}
\beq \Delta\f=\div(\F_x\gps)+\eps\Delta\vrf.   \eeqno{mailss}
Equation (\ref{iiatr}) is solved for $\psi$, which is then substituted in
Eq. (\ref{vvvfss}), to get an AQUAL-type equation of the form $\div[\tilde\m(|\gvrf|,\vr)\gvrf]=-\fpg\eps\r$
(with the added dependence of $\tilde\m$ on position, which does not complicate the solution). Substituting the solutions for $\psi$ and $\vrf$ in the right-hand side of Eq. (\ref{mailss}) gives the Poisson equation for the MOND potential.
\par
In this case, the ellipticity condition for Eq. (\ref{vvvfss}) can be shown to be that $y\F_y$ is a monotonic function of $y$

\subsection{AQUAL and QUMOND as special cases}
AQUAL is a special case of the tripotential class, gotten by putting $\F_x=0$, $\F_z=\eps$ a constant, so $\F=\eps z+\tilde\F(y)$, with $\tilde\F(y)$ having the appropriate limits. With these choices we have $\f=\eps\vrf$, while $\vrf$ satisfies an AQUAL equation
\beq \div(\tilde\F'\gvrf)=-\fpg\eps\r.   \eeqno{vvvflur}
In the high-acceleration limit $\tilde\F'\rar -\eps\^2$, and for small values of the argument $\tilde\F\rar -(2/3)\eps\^3y\^{3/2}$.
\par
QUMOND is a special case gotten for the choice $\F_y=\F_z=0$, and $\F(x)$ having the appropriate low- and high-acceleration limits. For these, Eq. (\ref{vvvf}) is satisfied trivially, and Eq. (\ref{mail}) becomes
\beq \Delta\f=\div(\F_x\gps)\equiv\fpg\hat\r,   \eeqno{mailqum}
which is the QUMOND field equation.\footnote{Remember that the ellipticity condition for AQUAL is that $X\m(X)$ be monotonic, and the QUMOND equations, which require solving only the Poisson equation, are elliptic.}
\subsection{Newtonian limit  \label{newt}}
To recover the Newtonian limit when $\az\rar 0$ -- one of the basic MOND axioms -- we require that in this limit $\F_z\rar \eps$, $\F_y\rar\eta$, and $\F_x\rar\b$, all constants, with $\b-\eps^2/\eta=1$. This can be seen from the field equations, or by noting that with these choices
the Lagrangian can be written as
\beq \L=-\frac{1}{\epg}[(\gf)^2-(\grad\t)^2-(\grad \z)^2
]  +\r(\oot\vv^2-\phi),  \eeqno{ladyra}
where, $\t=\f-\psi$, and $\z=\eps\eta\^{-1/2}\psi+\eta\^{1/2}\vrf$, which is the Poisson Lagrangian for $\f$ (and implying the irrelevant $\psi=\f$, and $\z=0$).
\subsection{The deep-MOND limit   \label{dml}}
Another basic MOND tenet is that MOND gravity becomes space-time scale invariant in the limit $\az\rar\infty$ (while $G\rar 0$, with $\A_0=G\az$ fixed) \cite{milgrom09a}.
\par
In general, scale invariance of a theory means that we can assign to each degree of freedom, and each independent variable -- such as time and positions -- $\U_i$, a scaling dimension $\a_i$, such that the theory is invariant under $\U_i\rar\l\^{\a_i}\U_i$. (The scaling dimension in not necessarily related to the $[\ell][t][m]$ dimensions.)
\par
For MOND, we specifically dictate that the DML is space-time scale invariant, namely, that the scaling dimensions of length and time are the same (and can be taken as $1$). It is also dictated that masses do not scale (i.e., have scaling dimension zero).\footnote{When one speaks of scale invariance in the context of quantum theories, one requires, in contradistinction to MOND, that masses have scaling dimension $-1$. This has to do with dimensions of the Planck constant, which appears in quantum theories.} So, in our context, a scaling transformation is $(t,\vr,m)\rar (\l t,\l\vr,m)$, and to ensure scale invariance, all terms in the DML of the Lagrangian have to transform in the same way under such scaling.
\par
The last term of Eq. (\ref{lagran1}) then dictates that the scaling dimension of $\f$ is zero, since velocities do not scale. Since masses do not scale, the scaling dimension of the baryon density, $\r$, has to be $-3$. This then dictates that the scaling dimensions of all the terms in the DML of $\L$ have to be $-3$. The first term in $\L\_G$ cannot be canceled by terms in $\F$; so, its having scaling dimensions $-3$ dictates that $\psi$ has scaling dimensions $-1$ (which should, of course, agree with $\psi$ being the Newtonian potential, and scaling as $m/r$).
\par
Now, $\vrf$ is the only remaining degree of freedom, whose scaling dimension is not dictated by the MOND axioms and the explicit terms in the TRIMOND Lagrangian. To ensure MOND scale invariance of TRIMOND, the DML of $\F$, call it $\F\_D$, must also have scaling dimension $-3$ for some assignment of scaling dimension, $\a$, to $\vrf$. This $\a$ depends on the theory; examples of how it is determined are given in Sec. \ref{oneddml}.
\par
Given this $\a$ and with the above scaling dimensions, the arguments of $\F$ scale as $x,y,z\rar\l^{-4}x,\l^{2(\a-1)}y,\l^{\a-3}z$.
Thus, to ensure MOND scale invariance, $\F\_D$ should be such that there is a choice of $\a$ for which
\beq \F\_D(\lambda^{-4}x,\lambda^{2\a-2}y,\lambda^{\a-3}z)=\lambda^{-3}\F\_D(x,y,z).   \eeqno{siff}
By taking $\l=x^{1/4}$, we see that $\F\_D$ can then be written as
\beq \F\_D(x,y,z)=x^{3/4}\F\_D[1,yx^{(\a-1)/2},zx^{(\a-3)/4}]\equiv x^{3/4}\bar\F\_D[yx^{(\a-1)/2},zx^{(\a-3)/4}].   \eeqno{niuter}
Such functions are the most general that ensure scale invariance.\footnote{When we consider the scaling transformation $\U_i\rar\l\^{\a_i}\U_i$, $\l$ must be taken as a constant. Otherwise, with $\l$ being, for example, position dependent, $\grad\U_i$ would not transform simply by scaling, and, in particular, would not simply have a scaling dimension $\a_i-1$, as we take when deducing the scaling dimensions of $x,~y,~z$.
This, however, does not prevent us from taking $\l=x^{1/4}$, which {\it does not mean} that we are employing some scaling transformation with space-dependent scale factor. Once the scaling dimensions of the variables $x,~y,~z$ themselves have been established (under transformations with constant scaling factors), Eq. (\ref{siff}) becomes an algebraic (i.e., not involving derivatives of the variables), necessary and sufficient condition for scale invariance. It should hold for any choice of the four {\it numbers} $\l,~x,~y,~z$; in particular for the choice  $\l=x^{1/4}$. Equation (\ref{niuter}) can also be deduced as follows: Since $\F\_D$ has to scale as $\l^{-3}$, and since $x^{3/4}$ scales like $\l^{-3}$, $\F\_D/x^{3/4}$ must be scale invariant, and must thus be a function of only scale-invariant quantities constructed from $x,~y,~z$. The only independent such quantities are $yx^{(\a-1)/2}$, and $zx^{(\a-3)/4}$.}
\par
So, unlike AQUAL or QUMOND where there is no freedom do dictate an interpolating function in the deep-MOND limit, here the DML is described, generally, by a function of two variables. These variables, and hence $\bar\F\_D$, are scale invariant, and the $ x^{3/4}$ prefactor gives $\F\_D$
a scaling dimension $-3$, as required.
\section{Some consequences and predictions   \label{conseq}}
\subsection{One-dimensional configurations  \label{oned}}
As in AQUAL and QUMOND, for systems of one-dimensional symmetry (planar, cylindrical, or spherical) the MOND acceleration $g$ is a system-independent (but theory specific) function of the Newtonian acceleration, $\gN$:
\beq  g=\n(\gN/\az)\gN.  \eeqno{lioper}
To see this, note that in this one-dimensional case we can strip the divergences in the field Eqs. (\ref{vvvf}) and (\ref{mail}) (e.g., by applying Gauss's theorem) and we can replace the vectorial relations by scalar ones with the absolute values of the respective accelerations replacing the potential gradient: $\gps=\gN\ve,~\gvrf=g\_\vrf\ve,~\gf=g\ve$ (where $\ve$ is the radial unit vector). Equation (\ref{vvvf}) than reads (putting $\az=1$ for the nonce)
\beq  \F_y(\gN\^2,g\_\vrf\^2,2\gN g\_\vrf)g\_\vrf+\F_z(\gN\^2,g\_\vrf\^2,2\gN g\_\vrf)\gN=0,   \eeqno{mumuiop}
from which $g\_\vrf$ can be solved for as a function of $\gN$. $\F$ has to be such that the solution for $g\_\vrf$ exists and is unique (the ellipticity of Eq. (\ref{vvvf}) should ensure this).
Equation (\ref{mail}) gives, in the same vein,
\beq  g=\F_x(\gN\^2,g\_\vrf\^2,2\gN g\_\vrf)\gN+\F_z(\gN\^2,g\_\vrf\^2,2\gN g\_\vrf)g\_\vrf.    \eeqno{mumulig}
Substituting the $g\_\vrf$ values gotten from Eq. (\ref{mumuiop}), we then get an equation of the form (\ref{lioper}).
\par
The function $\nu$ is an example of an emerging interpolating function. For the general TRIMOND theories it is relevant only for one-dimensional configurations. It can be used as some tool for comparison with AQUAL and QUMOND, where it is the only such function that appears and that is relevant to all phenomena and configurations.
\par
In terms of
\beq  \F^*(\gN,g\_\vrf)\equiv\F(\gN\^2,g\_\vrf\^2,2\gN g\_\vrf),  \eeqno{nanata}
 Eqs. (\ref{mumuiop}) and (\ref{mumulig}) can be written as
\beq  \frac{\partial\F^*}{\partial g\_\vrf}=0,  \eeqno{rebute}
\beq  g=\oot\frac{\partial\F^*}{\partial \gN}.  \eeqno{ropye}
\par
For nonsymmetric systems -- e.g., in predicting the rotation curves of disc galaxies -- TRIMOND theories do differ from each other. In fact, AQUAL and QUMOND themselves differ in such predictions (e.g., Ref. \cite{chae22}). But the fact that general TRIMOND theories revolve on Lagrangian functions of three variables can increase the variety in predictions.
To bring out better the possible variety in behaviors for nonsymmetric systems we can work with the acceleration variable $w\equiv(xy)-z\^2/4$ instead of with $z$ itself.
Since $w=0$ for one-dimensional configurations, the one-dimensional ``interpolating function,'' $\n$, is oblivious to the dependence of $\F$ on $w$. See more on this in Sec. \ref{oneddml}.
\subsubsection{The DML of one-dimensional configurations  \label{oneddml}}
For DML, one-dimensional configurations, we can further write, using the DML form, Eq. (\ref{niuter}),
\beq  \F^*(\gN,g\_\vrf)=\gN\^{3/2}\bar\F\_D[\gN\^{\a-1}g\_\vrf\^2,2\gN\^{(\a-1)/2}g\_\vrf]\equiv \gN\^{3/2}f(q),  \eeqno{nanare}
where $q\equiv g\_\vrf \gN\^{(\a-1)/2}$.
Equations (\ref{rebute}) and (\ref{ropye}) then become, respectively,
\beq f'(q)=0, ~~~~~g=\frac{3}{4}\gN\^{1/2}f(q).\eeqno{masutwe}
The first of Eqs. (\ref{masutwe}) tells us that its solution $q=q\_0$ is a universal constant of the theory.
The ellipticity condition of the theory should imply, in particular, that this equation has a unique solution. Then,
\beq g\_\vrf=q\_0\az(\gN/\az)\^{(1-\a)/2},~~~~~~ g=\frac{3}{4}f(q\_0)\sqrt{\gN\az}, \eeqno{momate}
 for all one-dimensional, DML configurations (where I reinstated $\az$).
\par
We see that the DML (and, in fact, the general) dependence of $g\_\vrf$ on $\gN$ in the one-dimensional case, is theory dependent, but scale invariance dictates that $g\propto\sqrt{\gN\az}$ in any MOND theory.
\par
It is customary to take the normalization of $\az$ so that in the DML of one-dimensional systems we have $g=\sqrt{\gN\az}$; so, for example, the predicted mass-asymptotic-speed relation (which underlies the baryonic Tully-Fisher relation) takes the exact form $MG\az=V^4\_{\infty}$.
\par
Given the interaction Lagrangian function $\F$, satisfying the requirements of the Newtonian and the deep-MOND limits, this requirement determines the normalization of the DML behavior, by adjusting the normalization of $f(q)$ so that $f(q\_0)=4/3$.
\par
For the sake of demonstration, consider a TRIMOND theory whose $\F$ has the DML form
\beq \F(x,y,z)~~~~ \overset{\az\rar \infty}{\longrightarrow}~~~~ \F\_D=\eps x\^\b z\^v+\tilde\F(y)+
\xi(4xy-z^2)^{\c/2}Q[yx^{(\a-1)/2},zx^{(\a-3)/4}],  \eeqno{lalayure}
where $\a$ is the scaling dimension of $\vrf$.
The scaling dimension of the first term having to be $-3$, implies that we have to take $\a=-3(1-v-4\b/3)/v$.  The scaling dimension of $y$ is $-2(1-\a)$; so we have to have $\tilde\F(y)=-sy\^{3/2(1-\a)}$.
\par
Also, we need to take $\c=3/(3-\a)$, so the last term has a scaling dimension $-3$, and can be written in the standard DML form, Eq. (\ref{niuter}): $\xi x\^{3/4}\{4yx^{(\a-1)/2}-[zx^{(\a-3)/4}]^2\}^{\c/2}Q[yx^{(\a-1)/2},zx^{(\a-3)/4}]$, making $\F\_D$
scale invariant.
\par
For one-dimensional configurations, this last term vanishes. Proceeding then as discussed above,
one finds that
\beq  f(q)=2^v\eps q\^v-sq\^u, \eeqno{ququya}
where $u=3/(1-\a)=3v/(3-2v-4\b)$. Take $v$ and $\b$ such that $\a<1$ ($u>0$); thus $u>v$.
For $v\le 1$, the first of Eqs. (\ref{masutwe}) has a unique solution\footnote{Note that for $v>1$, we also have the unwanted solution $q=0$, which gives the solution $g=g\_\vrf=0$; so such a theory does not make sense. We thus have to take $v\le 1$.}
\beq q\_0=\left(\frac{\eps v 2^v}{su}\right)^{1/(u-v)}.  \eeqno{numanu}
 The condition $f(q\_0)=4/3$ then fixes $s$ in terms of $\eps$.
\par
For the special case of AQUAL, $\F$ itself has the form (\ref{lalayure}), with $v=1$, $\xi=0$, and $\b=0$. Thus, $f(q)=2\eps/q-sq\^3$, $q\_0=(2\eps/3s)\^{1/2}$, and $s=2\eps\^3/3$.
\par
For any isolated system of bounded mass distribution, the asymptotic region, far outside the mass, is described by a spherical, DML solution; so the above applies to it. In particulary, we have asymptotically $g\rar (MG\az)^{1/2}/r$.

\par
We see that we can have different theories, with different interpolating functions, all being scale invariant in the DML, all having the same DML MOND accelerations (though not the same $\vrf$) in the one-dimensional case, but not for other configurations. For example, the predicted phenomenology of one-dimensional configurations, or of the asymptotic dynamics, does not inform us on the last term in Eq. (\ref{lalayure}). And, we know from our experience with numerical solutions of AQUAL, that, in general, asymmetric cases, $\gvrf$ and $\gps$ are not aligned. So, contributions to the Lagrangian, in which the last term in Eq. (\ref{lalayure}) appears, do not vanishes, and can be important in determining the MOND accelerations. It is difficult to make more concrete, quantitative statements without numerically solving various TRIMOND theories for various configurations.

\subsection{Deep-MOND virial relation and consequences}
The tripotential theories, with the MOND limit defined by Eq. (\ref{siff}), fall under the general class of MG, MOND theories dealt with in Ref. \cite{milgrom14b}. It was shown there that such theories generally satisfy a very useful virial relation: For a deep-MOND, self-gravitating, isolated system of pointlike masses, $m_p$, at positions $\vr_p$, subject to (gravitational) forces $\vF_p$, the following holds:
\beq \sum_p \vr_p\cdot\vF_p=-(2/3)(G\az)^{1/2}[(\sum_p m_p)^{3/2}-\sum_p m_p^{3/2}].  \eeqno{galama}
This then leads to the deep-MOND two-body force for arbitrary masses,
 \beq F(m_1,m_2,\ell)=\frac{2}{3}\frac{(\az G)^{1/2}}{\ell}[(m_1+m_2)^{3/2}-m_1^{3/2}-m_2^{3/2}] \eeqno{shasa}
($\ell$ is the distance between the masses),
and to a general mass-velocity-dispersion relation
\beq \s^2=\frac{2}{3}(MG\az)^{1/2}[1-\sum_p (m_p/M)^{3/2}],  \eeqno{shisa}
where $\s^2=M^{-1}\sum_p m_p\vv_p^2$ is the mass-weighted, three-dimensional, mean-squared velocity dispersion, and $M=\sum_p m_p$ is the total mass.
\par
This also implies that all these theories predict the same value of the $Q$ parameter, defined in Ref. \cite{milgrom12}, which was proposed as a possible discriminator between MOND theories. This parameter thus cannot discriminate between the different TRIMOND theories.
\par
The reason that the full detail of the DML of the tripotential theories does not enter the virial relation and its corollaries is that what effectively enters the derivation of this relation is only the asymptotic behavior of the MOND field from a single mass. And this is oblivious to the details of the theory,

\section{Discussion \label{discussion}}
The large variety afforded by the freedom to choose the three-variable $\F$ can potentially lead to a considerable variety in secondary predictions of TRIMOND theories. However, to demonstrate this in some quantitative detail would require numerical solutions, which I cannot provide here.
\par
There are several relativistic formulations of MOND that give
AQUAL and/or QUMOND as their nonrelativistic limits \cite{bekenstein04,zfs07,milgrom09b,babichev11,sanders11,marsat11,sz19,sz21,milgrom22}.
Some of these might be generalized, so as to result in a version of TRIMOND as their nonrelativistic limit. Note, however, that the general TRIMOND theory revolves around a function of three variables, while all the above theories employ functions of at most two variables.
Such generalizations would thus have to go beyond just adding scalar degrees of freedom, or by employing scalars that already emerge in their nonrelativistic limit.
\par
I exemplify this with the bimetric, relativistic formulation of MOND (BIMOND)
\cite{milgrom09b,milgrom22}, which can give either AQUAL or QUMOND as nonrelativistic limits, depending on the choice of the theory's parameters.
\par
It seems that generalizations of BIMOND with an additional, scalar degree of freedom could give TRIMOND theories as their Newtonian limits.
BIMOND theories use in their Lagrangians scalars that are quadratic in the tensors
\beq C\ten{\l}{\m\n}=\C\ten{\l}{\m\n}-\hat\C\ten{\l}{\m\n}, \eeqno{biba}
where $\C\ten{\l}{\m\n}$ and $\hat\C\ten{\l}{\m\n}$ are the connections of the two metrics that are the gravitational degrees of freedom in BIMOND. Such quadratic scalars reduce in the nonrelativistic limit to $(\grad\chi)^2$, for some potential $\chi$.
\par
With an additional, scalar degree of freedom, $\vrf$, we can construct additional scalar variables such as $\vrf\der{\m}\vrf\uder{\m}$ and the mixed $\vrf\der{\m}\bar C^{\m}$, $\vrf\der{\m} C^{\m}$, where $C^{\l}=\Gmn C\ten{\l}{\m\n}$, $ \bar C^{\l}=g^{\l\n}C\ten{\m}{\m\n}$ are the two traces of $C\ten{\l}{\m\n}$. Such variables are expected to reduce in the nonrelativistic limit to some $(\gvrf)^2$ and $\gvrf\cdot\grad\chi$.
So it might be possible to construct such scalar-bimetric theories that reduce to some TRIMOND theories.

%\clearpage
\end{document}